\documentclass{ws-ijgmmp}

\usepackage{mathtools}

\newcommand{\be}{\begin{equation}}
\newcommand{\ee}{\end{equation}}
\newcommand{\ba}{\begin{eqnarray}}
\newcommand{\ea}{\end{eqnarray}}

\newcommand{\lp}{\left(}
\newcommand{\rp}{\right)}
\newcommand{\lb}{\left[}
\newcommand{\rb}{\right]}

\newcommand{\0}{0$^\text{th}$}
\newcommand{\1}{1$^\text{st}$}
\newcommand{\2}{2$^\text{nd}$}
\newcommand{\3}{3$^\text{rd}$}

\newcommand{\diff}{\text{d}}

\begin{document}

\markboth{Tomi Koivisto}
{Energy in the Relativistic Theory of Gravity}

%
\catchline{}{}{}{}{}
%

\title{ENERGY IN THE RELATIVISTIC THEORY OF GRAVITY}

\author{TOMI KOIVISTO}

\address{Laboratory of Theoretical Physics, Institute of Physics, University of Tartu, W. Ostwaldi 1, 50411 Tartu, Estonia}
\address{National Institute of Chemical Physics and Biophysics, R\"avala pst. 10, 10143 Tallinn, Estonia}

\maketitle

\begin{history}
\received{31.12.2021}
\revised{(Day Month Year)}
\end{history}

\begin{abstract}

The more precise definition and the more fundamental understanding of the concepts of time, energy, entropy and information
are building upon the new, relativistic foundation of gravity. This lecture is an attempt to explain
the basic principles that underpin this progress, by focusing on the simple but subtle universal definition of energy. 
The principles are unearthed from Einstein's theory and Noether's theorems, beneath a century of misconceptions. 

\end{abstract}



\vskip25pt

Since Galileo, our understanding of motion rests upon two pillars, the principle of relativity and the principle of equivalence. The relativity principle states
that the laws of physics are the same in all inertial frames. The equivalence principle states the universality of gravitation. 
Einstein's insight was that movement in space and movement in time are connected by a new subprinciple of relativity, the invariance of the speed of light. An even greater theory, to encompass accelerated motion, was built upon the consolidated pillars \cite{einstein}. However, this structure was left unfinished and it remains incomplete, as we still lack the robust, universal answers to some quite basic questions of physics. For instance: what is energy?  

The prevailing view is that this is ``the wrong question'' \cite{Misner:1973prb}. According to the geometrical dogma, Einstein's theory is best paraphrased by the statement $G_{\mu\nu} = 8\pi G T_{\mu\nu}$. This does show that {\it matter is geometry}, but it tells little about gravity and little about inertia\footnote{See e.g. \cite{Lehmkuhl:2014qhu,Jimenez:2019woj,Lu:2021wif} for related discussions.}. The principle of equivalence is muddled. Making the statement with tensors renders it valid in 
arbitrary coordinate systems, but that is but a trivial realisation of the principle of relativity. It gives no 
distinction to inertial frames. Arguably, rather than generalised from the Lorentz group, the symmetry of relativity has degenerated to the unity group \cite{Fock:1957zz}.   
Surely, the search for a relativistic definition of energy is futile within the confines of the geometrical dogma.
What we shall pursue in this lecture is not to really contradict $G_{\mu\nu} = 8\pi G T_{\mu\nu}$, but to find its simpler expression that both reflects the Galilean principles of motion and actually concerns gravity. 

To discuss motion we first need introduce $\nabla_\mu$, the notion of a derivative.  
At the conference {\it Geometric Foundations of Gravity in Tartu 2017} we had proposed to incorporate the \1 principle of motion 
by formulating gravity as a translation gauge theory, and to incorporate the \2 principle of motion by postulating the integrability of that gauge theory \cite{Koivisto:2018aip}. Integrability simply means that the translation gauge covariant derivative can be defined by the property $[\nabla_\mu, \nabla_\nu] = 0$. In the following we shall refer to this derivative as {\bf the connection}.
We will indeed see that, besides {\bf the metric} $g_{\mu\nu}$, the independent connection $\nabla_\mu$ is instrumental \cite{BeltranJimenez:2017tkd,Jimenez:2022uvo} to the description of the basic properties of gravity. 


We should not confuse $\nabla_\mu$ with the $\prescript{g}{}{\nabla}_\mu$ induced by the metric, the unique symmetric connection that satisfies $\prescript{g}{}{\nabla}_\alpha g_{\mu\nu} = 0$.
The fact that test particles move along the geodesics of $\prescript{g}{}{\nabla}_\mu$
is a physical feature of Einstein's theory\footnote{Remarkably, this also is consistent
with the integrable translation gauge theory, since even though the Standard Model action is minimally coupled to the $\nabla_\mu$, the equations
of motion turn out to be minimally coupled to the $\prescript{g}{}{\nabla}_\mu$ (see lemma 1 and lemma 3 of ref.\cite{BeltranJimenez:2020sih}).};  but the fact that there always is a 
locally {\it freely-falling frame} wherein $\prescript{g}{}{\nabla}_\mu \overset{\text{f.f.}}{=} \partial_\mu$, is merely a tautology of differential geometry.
Any curve on a metric manifold is infinitesimally a straight line. Despite the ``devastating objections''  beginning from Einstein \cite{norton}, the existence of a freely falling frame is still often presented as an expression, even a defining property of the equivalence principle. 
    
On the other hand, the relativity principle presupposes the existence of {\it inertial frames} wherein the laws of physics assume their standard form. 
There is a crucial difference in the concept of an inertial frame and the concept of a freely-falling frame that we should now clarify. 
Whereas the freely-falling frame is a special coordinate system, an inertial reference frame should by definition be preserved by coordinate transformations. Only then it may accommodate the proper, generally covariant laws of physics. 
Therefore, the criterion that distinguishes inertial reference frames must involve both the metric and the connection, and the criterion must be invariant under simultaneous coordinate transformation of both the fields. Since the metric can thus be expressed in arbitrary coordinates, it is clear that in general an inertial frame need not be freely falling. 

Let us then proceed towards the precise definition of an inertial frame. It is useful to first introduce the kinetic gravitational field excitation tensor density $\mathfrak{h}^{\mu\nu}{}_{\alpha}=-\mathfrak{h}^{\nu\mu}{}_{\alpha}$, or 
more succinctly, {\bf the gravitational field strength} 
\be \label{field}
\mathfrak{h}^{\mu\nu}{}_\alpha = \frac{m_P^2}{\sqrt{-g}}g_{\alpha\gamma}\nabla_\beta\lp g g^{\beta[\mu}g^{\nu]\gamma}\rp\,,
\ee  
where $m_P = 1/\sqrt{8\pi G}$ is the Planck mass and $g$ is the determinant of the metric. The form of the gravitational field strength was derived in an axiomatic, premetric construction of the theory \cite{Koivisto:2019ggr}. It was also derived in the Noether formalism \cite{Jimenez:2021nup}.
The source of the gravitational field strength is the matter energy-momentum tensor density $ \mathfrak{T}^\mu{}_\nu = \sqrt{-g}T^\mu{}_\nu$ i.e. {\bf the material current}. 
Therefore, in an inertial frame the field equation for the gravitational field strength must assume the form
\be \label{ife}
\nabla_\alpha\mathfrak{h}^{\alpha\mu}{}_\nu\, \overset{\text{i.f.}}{=} \, \mathfrak{T}^\mu{}_\nu\,.
\ee
This resembles the inhomogeneous Maxwell equation, the only difference being an extra index which reflects the existence of four conserved (translation) currents instead of one conserved (phase) current. In this sense, the theory is an exact realisation of ``gravitoelectromagnetism''. At this point it is perhaps already dawning upon us that in the end,
the gravitational charges - energy and momenta - should not be so much more mysteriously elusive than the electromagnetic charges.  

The presence of the connection affords the existence of solutions which satisfy (\ref{ife}) for arbitrary metrics\footnote{A technical caveat is that insisting on a generic {\it global} inertial frame may require the extension \cite{BeltranJimenez:2019odq} to non-symmetric $\nabla_\mu$ which we can omit for the pedagogical purpose of this lecture. The covariance group would be extended from Diff$^2$ to GL$\times$Diff.}.  Another important role of the connection $\nabla_\mu$ is to support the covariance of the  gravitational field strength $\mathfrak{h}^{\mu\nu}{}_\alpha$. With the metric alone, we could not consistently capture the physics of gravity, even though the metric is sufficient for the description of dynamics. The dynamics is encoded into the equations of motion derived from the action, but besides the equations of motion, the variation of the action determines the invariants and the observables, by yielding the conserved currents of the theory \cite{noether}. Now there is a beautiful relation between these two aspects of the theory, the {\it dynamics} and the {\it currents}. With a generic connection, we can write the field equation as   
\be \label{ife2}
\underbrace{\nabla_\alpha\mathfrak{h}^{\alpha\mu}{}_\nu}_{\text{current from Noether's \2 theorem}} = \underbrace{\mathfrak{t}^\mu{}_\nu + \mathfrak{T}^\mu{}_\nu}_{\text{current from Noether's \1 theorem}}\,.
\ee
The tensor density $\mathfrak{t}^\mu{}_\nu$ is the energy-momentum tensor density of the metric field \cite{BeltranJimenez:2019bnx} i.e. {\bf the inertial current}. 
Comparison with (\ref{ife}) shows that we can regard the coordinate-independent condition $\mathfrak{t}^\mu{}_\nu = 0$ as the general-relativistic definition of {\bf an inertial frame}. 
The invariant Noether charges $E_\mu$ are computed as integrals over an arbitrary closed surface $s$ with the binormal $s_{\mu\nu}$,
\be \label{charge}
E_\alpha = \oint_s \mathfrak{h}^{\mu\nu}{}_\alpha s_{\mu\nu}\,,
\ee
and the physical result is obtained by evaluating this integral in an inertial frame. 
We have arrived at the universal definition of energy. Note that physical charges are fundamentally ``quasi-local'', which is as local as it gets since one can in principle consider an infinitesimal $s$. 

The theoretical structure we have uncovered gives a precise meaning to the principle of equivalence. We can always accelerate a reference frame by transforming $\nabla_\mu$ to generate a gravitational field strength with non-vanishing energy even when $g_{\mu\nu}=\eta_{\mu\nu}$. However, the physical energy is unique, and measured by an observer with the 4-velocity $u^\mu$ as $u^\mu E_\mu$. 
The metric field determines the results of measurements, yet any localisable energy-momentum $\mathfrak{t}^\mu{}_\nu$ associated with the metric field itself is unphysical. 
The connection also has a very special role and this field is adjusted, from the perspective of an inertial observer, such that $\mathfrak{t}^\mu{}_\nu=0$.
However, generalising the simplified field equation (\ref{ife}) to a generic (non-inertial) frame using (\ref{ife2}), we obtain $\mathfrak{T}^\mu{}_\nu =  \nabla_\alpha\mathfrak{h}^{\alpha\mu}{}_\nu - \mathfrak{t}^\mu{}_\nu = m_P^2\sqrt{-g}G^\mu{}_\nu$, where the connection is canceled from the right hand side. This way, we undisclose the principle of relativity and recover only the equivalence of matter and geometry. 

The rise of the geometrical dogma begun with the Hilbert's formulation of gravity, based on the geometric invariant $R= -G^\mu{}_\mu$, superseding Einstein's 
Hamiltonian formulation (see e.g. (47) of \cite{einstein}) even before its first publication. We have already emphasised the paramount importance of the action in determining the
two foundational aspects of the theory as expressed in (\ref{ife2}). Indeed, we should recognise the principle of least action as the \3 - or even better, the \0 - principle of motion. 
The original Hamiltonian was not invariant under general coordinate transformations, and it is there we can trace back the incompleteness of theory: the \0 principle of motion was not
quite compatible with the \1 principle. This is now fixed by taking into account the metric-independent connection $\nabla_\mu$ .   
Before illustrating (\ref{charge}) by computing an explicit example, let us point out that whilst its derivation required a refinement to the formulation of Einstein's 
theory, the relation (\ref{ife2}) should be considered as the canonical expression of Noether's theorems, involving no hidden ``improvements'' or ambiguities.
 
There had been many confusions surrounding both sides of the equation (\ref{ife2}) i.e. both of the Noether's two theorems. 
It was found only quite recently that the gravitational field strength (\ref{field}) is the Noether charge of the diffeomorphism gauge symmetry \cite{Jimenez:2021nup}. A key issue is the quasi-locality of the 
charges, reflecting the holographic property of the interaction. The surface integrals contain all the physical information, but they are often discarded by considering only asymptotic regions with suitable boundary conditions for the fields.  
Although it is known that the surface integrand a.k.a. the ``superpotential'' is fully determined by the boundary terms in variation of the action \cite{DeHaro:2021gdv}, it is often asserted that the superpotentials are ambiguous. However, that is the case only in the context of an incomplete theory wherein one may have to resort to different boundary terms for different purposes\footnote{In particular, this is the case for the Hilbert's action, which is sometimes necessary to modify by a non-covariant, higher-order boundary term, such as the Gibbons-Hawkins-York term. It is for the same reason that the Komar superpotential \cite{Komar:1958wp} fails in some situations. It will be shown elsewhere that the problem is alleviated by the generalisation of the Komar
superpotential in the Einstein-Palatini formulation.}. With a well-defined action principle, the currents are no more ambiguous than the 
field equations. 

There are common misconceptions even about the material current $\mathfrak{T}^\mu{}_\nu$. 
A non-covariant method leads to a {\it pseudo-canonical pseudo-tensor}, and then some ad hoc ``improvement'' is used to modify it into a proper tensor. However, in the context of gauge theory, the canonical method is gauge-covariant and leads to the correct result. One obtains $\mathfrak{T}^\mu{}_\nu = \sqrt{-g}T^\mu{}_\nu$, where $T^\mu{}_\nu$ is the standard Hilbert energy-momentum tensor. Noether had already understood this, and the details were developed for the case of the electromagnetic field by Bessel-Hagen in 1921 \cite{Baker:2021hly}. For the case of the gravitational fields, the tensor density $\mathfrak{t}^\mu{}_\nu$ was introduced canonically (in particular, without resorting to ad hoc reference fields) as the metric Noether current in the translation gauge 
theory \cite{BeltranJimenez:2019bnx}. Fixing to the so called ``coincident gauge'' $\nabla_\mu \overset{\text{c.g.}}{=} \partial_\mu$ \cite{Jimenez:2022uvo} the inertial current reduces to the pseudo-canonical pseudo-tensor appearing in the original Hamiltonian formulation (see e.g. equation (50) in Ref. \cite{einstein}). 

The idea that the local energy-momentum current must be given solely by $\mathfrak{T}^\mu{}_\nu$ has been supported by various mathematical and physical arguments 
and is compatible with the alternative proposals for the definition of energy-momentum from Levi-Civita, Lorentz, Klein, etc \cite{book} to contemporary authors, e.g. \cite{Nikolic:2014kga,Aoki:2020prb}. In particular, Cooperstock's hypothesis was that since the energy-momentum conservation laws are devoid of content in the vacuum, the components of the Einstein's pseudo-tensor vanish in the appropriate coordinates \cite{cooperstock}. Therefore, we could call the definition of an inertial frame, $\mathfrak{t}^\mu{}_\nu = 0$, the {\it Cooperstock criterion}. Yet, this can be only a part of the story.  
The charge from Noether's \1 theorem would not allow to associate energy-momentum with gravitational waves nor with black holes without matter, but thanks to fantastic recent progress, there is now even experimental evidence for the reality of these gravitational phenomena. The resolution is the charge (\ref{charge}) from Noether's \2 theorem. But how can the surface reveal more than the bulk? 
The relation (\ref{ife2}) and the Gau{\ss} theorem seem to suggest the equivalence of the two charges. How can a definite integral over an everywhere-vanishing energy density give a non-vanishing energy?
The resolution is very simple, but its workings appear to be highly non-trivial!

 
Let us illustrate this in the case of a black hole, where topology has a part in the non-trivial play. To take into account both metric and matter fields, we consider the Reissner-Nordstr\"om-de Sitter black hole described by
\ba
\diff s^2 & = & -f(r)\diff t^2 + \diff r^2/f(r) + r^2\diff \Omega^2\,, \nonumber  \newline \\ \text{where} \quad f(r) & = & 1-\frac{m_S}{4\pi m_P^2r} - \frac{q^2}{r^2} + \frac{\Lambda}{3}r^2\,. 
\ea  
This solution requires a cosmological constant $\Lambda$ and an electric field associated with the charge $q$, whereas $m_S$ is the Schwarzschild mass parameter. To set the solution in the form (\ref{ife}),
we also need a non-vanishing connection. Considering then the energy contained within a radius $r$ from the singularity, we obtain from (\ref{charge}) the expression
\be
E_0(r) = 4\pi m_P^2\lb 1-f(r)\rb r = m_S - \frac{4\pi m_P^2 q^2}{r} + \frac{4}{3}\pi r^3\rho_\Lambda\,. 
\ee
Note that also the energy contributed by the sources (the $\Lambda$ and the electric field) is computed from their impact to the gravitational field strength and not directly from the sources themselves. Nevertheless, we obtain
precisely the expected Coulomb potential energy due to the charge $q$, and the correct volume integral of the constant energy density $\rho_\Lambda = m_P^2\Lambda$. 
To understand the contribution surviving even in vacuum, we note that the relevant component of the gravitational field strength (\ref{field}) turns out to have the simple form
$\mathfrak{h}^{0r}{}_0 = m_P^2(1-f)/r = m_S/4\pi r^2$, where the last equality is for the pure Schwarzschild case. The gravitational field has the Newtonian $1/r^2$ behaviour, precisely like the static electric field due to a point charge. The gravitational field force lines would terminate at the origin, and thus the divergence of the field strength there would be non-zero - but now the point at the origin does not exist in the manifold.

\section*{Acknowledgments}

The author acknowledges relativistic and energetic discussions with the organisers and participants of the conference {\it Geometric Foundations of Gravity in Tartu 2021}. This work was supported by the Estonian Research Council grants PRG356 ``Gauge Gravity'' and MOBTT86, and by the European Regional Development Fund CoE program TK133 ``The Dark Side of the Universe''.


\end{document}